# A natural mechanism for l-homochiralization of prebiotic aminoacids

Falcon F.L.

*Materials Science and Technology Institute, University of Havana, Cuba*

**Abstract**

We propose a mechanism that explains in a simple and natural form the l-homochiralization of prebiotic aminoacids in a volume of water where a geothermal gradient exists.

This model demands the following three assumptions:

1- The process takes place in volume of water where there is a vertical thermal gradient (Fig.1). This situation is common even nowadays. For example, the subglacial lakeVostok, in Antarctica, maintains a water temperature of T ≈ 3 $^0$C throughout the year, due to geothermal causes.
2- The relative composition of aminoacids dissolved in the volume of water is very close the one obtained in the Miller-Urey experiments.
3- The concentration of the dissolved aminoacids increases very slowly, beginning from a level where its precipitation doesn't occur yet.

Assumptions 2 and 3 can be naturally achieved taking into account that, during the Azoic Age, Earth's surface water reservoirs were for a long time doped with the same aminoacids composition due to Miller-Urey atmospheric reaction.

On the other hand, the slow increase of aminoacids concentration can be accomplished by the cooling that Earth's suffered at the end of the Azoic Age, which caused the partial freezing of liquid water in surface water reservoirs. The decreasing of liquid water volume and the segregation of amino acid molecules by the solidified water (ice) led to the slow increase of concentration of dissolved aminoacids in these water reservoirs. This concentration increase could reach such levels that crystallization may have taken place.

This supposition is in full accordance with the nowadays growing opinion that life could be originated under freezing conditions [1].

When we relate the relative concentration and the thermal dependence of solubility of each aminoacid obtained in Miller-Urey experiment [2, 3], we conclude that alanine must be the first substance that precipitates when the overall aminoacids concentration slowly increases.

Fig.2 shows the thermal dependence of *l* and *dl*-alanine. In our knowledge, there isn't any published work where the solubility difference between *d* and *l*-alanine was measured. For this reason, we calculated and compared both solubilities, considering the equation [2]:

$$\ln K_s = a + b\,T + c\,\ln T$$

where:

$K_s$ – thermodynamic solubility constant

T – temperature ($^0$K)

$a$, $b$ and $c$ – specific constants for each substance:

| amino acid | a | b | c |
|---|---|---|---|
| *d*-alanine | -41.6326 | 919.4029 | 6.1779 |
| *l*-alanine | -41.6730 | 920.6575 | 6.1843 |

Evaluating $K$(T) for *d* and *l*-alanine, we don't observe any difference in the range 273-300 $^0$K and a slow increase of *l*-alanine over *d*-alanine solubility in the range 300-373 $^0$K, reaching a maximum difference in their solubility constants of 0.08% at 373 $^0$K.

Fig.3 shows a section of Fig.2 where we can observe that when alanine concentration exceeds the value **C**, corresponding to equilibrium at the maximal temperature of system (**T$_{max}$**), will begin the nucleation of pure *l* and *d*- alanine crystals or conglomerates, because *dl*-alanine has a greater solubility.

In [4] are shown the main crystallographic data of *d* and *l*-alanine at 270 K:

|  | *d*-alanine | | *l*-alanine |
|---|---|---|---|
| Crystal system | | Orthorhombic | |
| Space group | | P212121 | |
| Unit cell dimensions | | | |
| a (Å) | 6.0073 | | 6.0095 |
| b (Å) | 12.3030 | | 12.3388 |
| c (Å) | 5.7732 | | 5.7904 |
| α | | 90$^0$ | |

| | β | 90⁰ | |
| --- | --- | --- | --- |
| | γ | 90⁰ | |
| | Volume (Å³) | 426.69 | 429.36 |

As we can see, at ice-water equilibrium temperature the unitary cell of *l*-alanine has greater volume than *d*-alanine cell. A possible explanation of this difference would be the energy difference due to parity violation (EDPV), that causes the weakness of intermolecular bonds due to strengthen of intramolecular ones. The supposition presented before needs to be confirmed, but in the case that the unitary cell volume difference was due to other causes (e.g. the different capacity of *l* and *d*-alanine to occlude disordered water into the crystal) the soundness of this model would be maintained.

As a consequence of before, a crystal or conglomerate of *l*-ala, with the same number of molecules than other one of *d*-ala, will have greater size, and indeed, greater effective cross-section when it is dragged by a convective flow.

Fig.4 shows what must happen in our system when alanine concentration slowly exceeds concentration **C.** Due to vertical thermal gradient, will arise convective flows that will drag the still little *l* and *d*-nuclei, following the arrows in Fig. As these nuclei go over different temperatures along their paths, they will partially dissolve when they pass closer to warmer bottom and will grow during their ascension to colder top of the pool. The maximum height that they can reach will depend of how much they will grow during ascension. When their apparent weight in solution is equal to ascending drag force, obeying Stokes' Law, they reach their maximal height and begin their descent. Because the *l*-nuclei have greater cross-section than *d*-nuclei, the ascending drag force on they will be greater and, as a consequence, they will reach greater height than *d*-nuclei before their apparent weight become equal to ascending drag force and begin their descent. By this reason, the *l*-nuclei can reach a colder zone in the pool, which *d*-nuclei can't reach. In consequence, the *l*-nuclei undergo a greater increase of size than *d*-nuclei in each ascent-descent cycle.

This process is shown in Fig.4. The *l*-nuclei reach a radius $r_{l\,max}$ greater than $r_{d\,max}$, because *l*-nuclei reach isotherm **T₂** which is colder than **T₁**, which is the coldest isotherm that can reach *d*-nuclei, because their higher density.

Because the solubilities of *l* and *d*-nuclei are very close, their volume difference practically maintains when they partially dissolve during their pass through the warmer zone near the bottom of the pool ($r_{l\,min} > r_{d\,min}$) and will slowly increment along cycles repetition, during the slow rise of aminoacids concentration in the system.

In addition to before exposed, the dissolved in solution aminoacids will suffer two processes that will tend toward the maintenance of their racemic composition:

1. Their individual racemization, a very slow process in alanine but, as an energy activated process, can be enhanced during the pass of these molecules through the warmer zone in the bottom of the pool.
2. The induced epimerization, that is, the change of chirality of *d*-molecules when they are added to *l*-crystals or conglomerates, in order to reach a more stable energetic state.

Brief calculations:

A *l*-nucleus with the same number of molecules than a *d*-nucleus, will have a greater radius. This radius relation is the same that of their unitary cells:

$$\frac{r_L}{r_D} = 1,002 \qquad (1)$$

The relation **r**$_{l\,max}$ / **r**$_{d\,max}$ can be calculated equalizing the ascending drag force expressed in Stokes' Law and the apparent weight of nucleus in solution:

$$6\pi r \eta v = \frac{4}{3}\pi r^3 (\rho_p - \rho_{sol})g \qquad (2)$$

Where:
$r$- nucleus radius
$\eta$- solution viscosity
$v$- nucleus velocity
$\rho_p$- nucleus density
$\rho_{sol}$- solution density
$g$- gravity acceleration

When *l* and *d*-nuclei reach the top of their paths, their vertical velocities are the same and equal to null. Evaluating equation (2) for *l*- and *d*-nuclei and equalizing their velocities we obtain the relation:

$$\frac{r_{l\,max}^2}{r_{d\,max}^2} = \frac{\rho_d - \rho_{sol}}{\rho_l - \rho_{sol}} \qquad (3)$$

Where $\rho_d$ and $\rho_l$ are the densities of *d* and *l*-nuclei, respectively.

The magnitude $\rho_{sol}$ is unknown. For this reason we calculated it adding the volumes of water and dissolved alanine, without considering dissolution contraction and the increase of $\rho_{sol}$ due to other dissolved in solution molecules, which results in the minimal value of relation (3):

$$\frac{r_{l\,max}^2}{r_{d\,max}^2} > 1,030$$

which implies:

$$\frac{r_{l\,max}}{r_{d\,max}} > 1,015$$

as can be seen, this radii relation between *l* and *d*-nuclei exceeds relation (1), with the conclusion that *l*-nuclei experiment greater growth than *d*-nuclei during their ascension.

As follows, after nuclei grown in colder zone, they descent to the warmer bottom, partially dissolve and their volumes diminish. The volumes relation between *l* and *d*-nuclei will be affected according their solubility relations:

$$\frac{r^3_{l\,min}}{r^3_{d\,min}} = \frac{r^3_{l\,max}}{r^3_{d\,max}} \cdot \frac{1}{1,0008} = \mathbf{1,045}$$

From which,

$$\frac{r_{l\,min}}{r_{d\,min}} = \mathbf{1,0147 \approx 1,015}$$

This result says that *l* and *d*-nuclei maintain nearly constant their size relation during their dissolution near the bottom of the pool.

In summary, during ascension *l*-nuclei suffer greater growth than *d*-nuclei, because they reach colder zones in the pool due to their lower density. During descent, this size difference maintains, because their minimal solubilities difference. Adding to before, the epimerization process, which will convert dissolved *d*-aminoacids to *l*, results in a natural Ostwald ripening process which leads the totality of aminoacids in the system to their complete *l*-homochiralization.

**Acknowledgments**

We thank C.M. Falcon for helpful discussions.

**Figure captions**

Fig.1-Diagram of isotherms in a water pool with vertical thermal gradient.

Fig.2-Thermal dependence of solubility for *dl* and *l*-alanine

Fig.3- Section of Fig.2, where **C-** equilibrium concentration of *d* and *l*-alanine at **T**$_{\mathbf{max}}$ (ground temperature in a geothermal water pool)

Fig.4- Convective trajectories and radii of *d* and *l*-nuclei in system of Fig.1, where **r**$_{d\,\mathbf{max}}$ and **r**$_{l\,\mathbf{max}}$ – maximal nuclei radii in the trajectories tops; **r**$_{d\,\mathbf{min}}$ and **r**$_{l\,\mathbf{min}}$ – minimal nuclei radii in the trajectories bottom, for *d* and *l*-nuclei, respectively.

**Bibliography**

1- Alexander V. Vlassov, Sergei A. Kazakov, Brian H. Johnston, Laura F. Landweber; J Mol Evol (2005) 61:264–273
2- Xin Xu, Simao P. Pinho, Eugenia A. Macedo; *Ind. Eng. Chem. Res.* **2004,** *43,* 3200-3204
3- Dalton J. B., Schmidt C.L.A.; J. Biol. Chem., Vol. 103, No.2, 549-578
4- W.Q. Wang, Y. Gong, Z. Liang, F.L. Sun, D.X. Shi, H.J. Gao, X. Lin, P. Jiang, Z.M. Wang; Surface Science 512 (2002) L379–L384

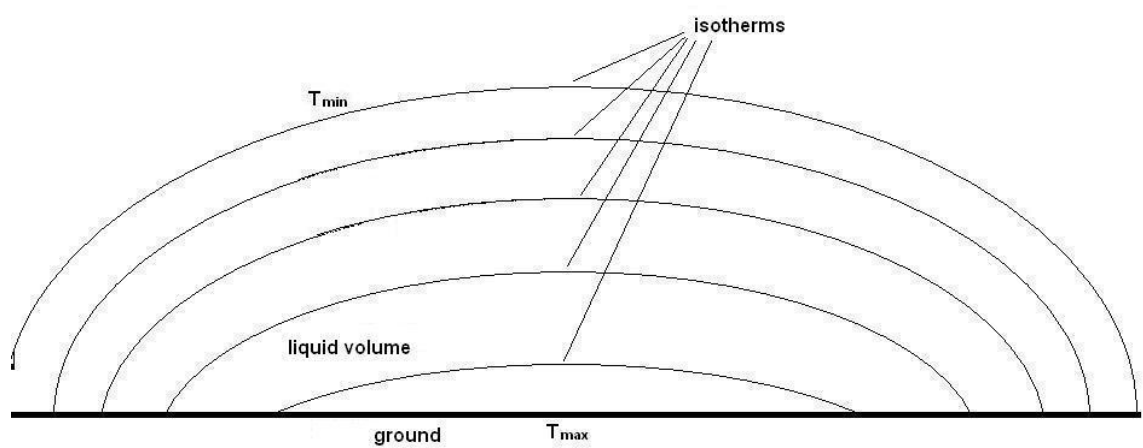

**Fig.1**

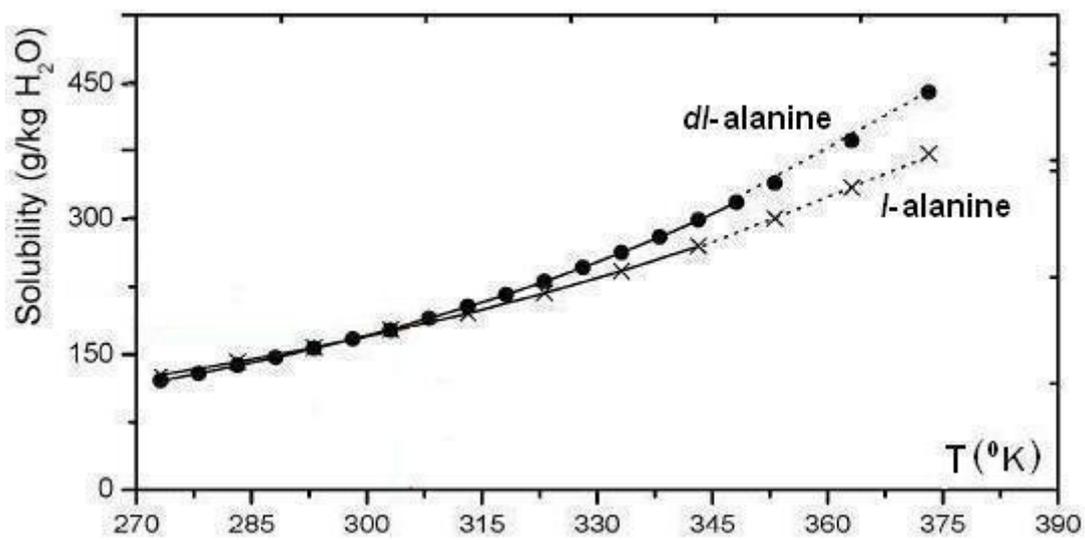

**Fig.2**

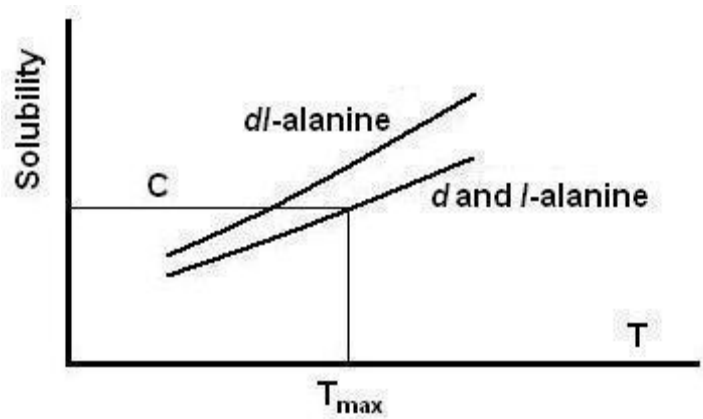

**Fig.3**

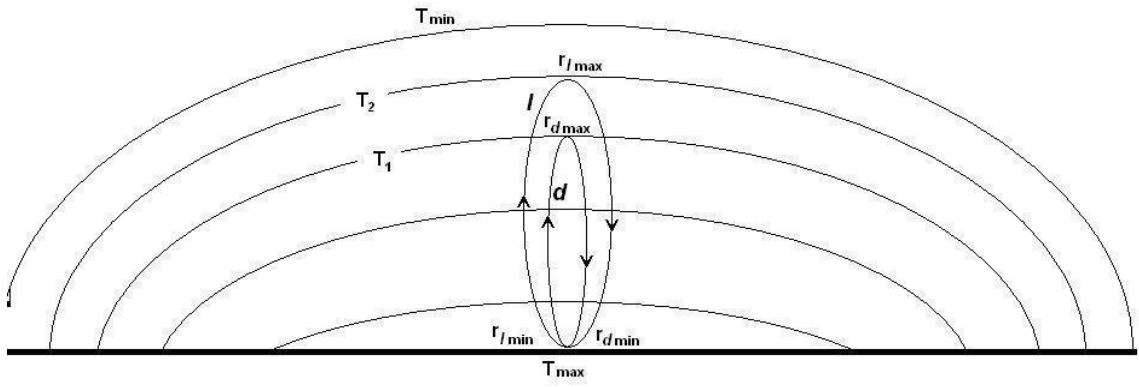

**Fig.4**